\documentclass[12pt]{article}
\usepackage{amsmath, amssymb, amsthm, fullpage, parskip, setspace}
\usepackage[numbers, sort&compress]{natbib}

\usepackage{algpseudocode}
\usepackage{graphics, graphicx}
\usepackage{epsf}
\usepackage{float, multirow,parskip, subcaption, setspace}
\usepackage{comment}
\usepackage{url}
\usepackage{enumitem}
\usepackage{tikz}
\usepackage{bm, bbm}
\usetikzlibrary{shapes.geometric, positioning}
\usetikzlibrary{quotes, angles}
\usepackage{rotating}
\usepackage{xr, xr-hyper}
\usepackage{hyperref}
\usepackage{placeins}
\usepackage{algorithm2e}
\usepackage{xspace}
\usepackage{xcolor,soul}
\definecolor{vlgray}{gray}{0.9}
\sethlcolor{vlgray}

\usetikzlibrary{shapes}

\RestyleAlgo{ruled}

\definecolor{SkyBlue}{RGB}{14, 118, 188}
\definecolor{BrightRed}{RGB}{223,82, 78}

\hypersetup{pdfborder = {0 0 0.5 [3 3]}, colorlinks = true, linkcolor = BrightRed, citecolor = SkyBlue}

\theoremstyle{plain}

\theoremstyle{definition}

\theoremstyle{Remark}

\title{Adolescent sports participation and health in early adulthood: An observational study}
\author{Ajinkya H.~Kokandakar\thanks{University of Wisconsin--Madison}, Yuzhou Lin\thanks{Harvard University}, Steven Jin\thanks{University of Pennsylvania}, \\Jordan Weiss\thanks{Stanford University}, Amanda R.~Rabinowitz\thanks{Moss Rehabilitation and Research Institute}, Reuben A. Buford May\thanks{University of Illinois Urbana-Champaign}, \\ Dylan Small\footnotemark[3], and Sameer K.~Deshpande\footnotemark[1]}

\begin{document}
\maketitle

\begin{abstract}
We study the impact of teenage sports participation on early-adulthood health using longitudinal data from the National Study of Youth and Religion.
We focus on two primary outcomes measured at ages 23--28 --- self-rated health and total score on the PHQ9 Patient Depression Questionnaire --- and control for several potential confounders related to demographics and family socioeconomic status.
To probe the possibility that certain types of sports participation may have larger effects on health than others, we conduct a matched observational study at each level within a hierarchy of exposures.
Our hierarchy ranges from broadly defined exposures (e.g., participation in any organized after-school activity) to narrow (e.g., participation in collision sports). 
We deployed an ordered testing approach that exploits the hierarchical relationships between our exposure definitions to perform our analyses while maintaining a fixed family-wise error rate.
Compared to teenagers who did not participate in any after-school activities, those who participated in sports had statistically significantly better self-rated and mental health outcomes in early adulthood.

\end{abstract}

\section{Introduction}
\label{sec:introduction}
More than 6 million high school students in the United States of America participate in organized sports every year \citep{nhfs_2021_22}.
The prevalence of sports participation --- estimated to account for 50-60\% of the high school population over the past two decades  \citep{cdc_school_participation} --- reflects widespread interest in athletics and underscores the integral role that sports play in the educational and social development of American high school students.
Prior work has established positive associations between high school sports participation and physical and mental health as well as life satisfaction \citep{Zarrett2018,Donaldson2006,zullig_white_2011,Jewett2014,song_shi_2023}.

While evidence of the positive effects of sports participation is voluminous, risk of sports-related injury may deter some children and families from involvement \citep{fishman_et_al_2017,hibshman_et_al_2022}. 
Collision sports, in particular, have faced heightened scrutiny due to perceived risks of chronic traumatic encephalopathy (CTE), a neurodegenerative disease \citep{mez_et_al_2017} that has received attention in both the scientific literature \citep{mez_et_al_2017,deshpande_et_al_2017} and popular media \citep{lannan_gbh,omalu_nyt,nowinski_vox,gregory_time,arnold_usa_today,nyt_nfl,landesman_movie,starapoli,schaller_philly_inq,chuck_nbc}. 
The potential long-term consequences of concussions must now be considered by families when making decisions about adolescent participation in sports. 
Indeed, there has been a decline in participation in organized sports among American high school students for the first time in 30 years, with the steepest decline in football, according to a 2019 report \citep{nhfs_2019}.

Decision-making about the relative benefits and risks associated with sports participation is complicated by a lack of consensus within the scientific community regarding the association between collision sports and neurodegenerative diseases \citep{wolfson_et_al_2020,stewart_et_al_2019,kuhn_et_al_2017}. 
A systematic review of long-term effects of sports-related concussion suggests that multiple concussions appear to be a risk factor for cognitive impairment and emotional problems in some people \citep{manley_et_al_2017}. 
Nevertheless, the challenge lies in discerning how families should interpret and incorporate this risk when contemplating sports participation for their child.
To date the majority of research on this topic has focused on either contemporaneous outcomes or later-life functioning, with relatively few studies examining the influence of sports participation in adolescence on physical and mental health outcomes in early adulthood.
Regarding later-life health outcomes, an emerging body of research suggests that, when examined collectively, individuals involved in contact sports fare no worse than their peers with regard to cognitive and emotional functioning later in life \citep{deshpande_et_al_2017,deshpande_et_al_2020,weiss_et_al_2021}.
A recent study \cite{bohr_et_al_2019} examined a subsample ($n=10{,}951$) from the National Longitudinal Study of Adolescent to Adult Health \citep[Add Health;][]{add_health_2013} to investigate the association between sports participation and various mental health outcomes in early adulthood.
Participants were categorized as participating in no sports, non-contact sports only, and contact sports. 
Multivariate and logistic regression models were employed, and the analysis was repeated on a males-only sample ($n = 5{,}008$). 
The authors found that intention to participate in contact sports was not significantly linked to any of the outcomes in the full-sample analysis.
However, the intention to participate in football was associated with reduced odds of depression diagnosis in adulthood compared to non-contact sports participation in the males-only sample. 
Notably, football showed no significant associations with impaired cognitive ability, increased depressive symptoms, or heightened suicide ideation \citep{bohr_et_al_2019}.

Amidst this backdrop, considerable uncertainty persists regarding the broader health impacts of sports participation as compared to other extracurricular activities.
Motivated by this uncertainty, we use data from the National Study of Youth and Religion \citep[NSYR;][]{NSYR_design} to estimate the effect of multiple types of sports participation on physical and mental health in early-adulthood.
Specifically, we not only estimate the broad effect of playing any sports but also estimate the more granular effects of playing non-contact (e.g., track, swimming, and golf), non-collision contact (e.g., basketball, soccer, and volleyball), and collision sports (e.g., football, wrestling, and hockey).
Although our approach involves testing several hypotheses, we are nevertheless able to control the family-wise error rate (FWER) through a testing-in-order approach \citep{Rosenbaum_testing} that exploits the logical relationships between different types of participation.
We find that the subjects in our sample who take part in any after school activity are more likely to self-report as healthy, experience fewer depressive symptoms, and have higher life satisfaction than comparable subjects who do not participate in any activity.
Subjects who participated in contact sports were significantly more likely to self-report as healthy than comparable subjects who did not participate in any extracurricular activities.
Additionally, subjects who participated in non-collision contact sports experienced significantly fewer depressive symptoms than comparable subjects who did not participate in any activities.
Finally, we find that the subjects who participated in collision sports were not statistically significantly different from comparable control subjects in terms of self-reported health and depression scores.

\section{Data and Methods}
\label{sec:methods}
Before analyzing the outcomes, we published our pre-analysis protocol \cite{kokandakar2024preanalysis}, which contains further details about our analysis sample, our matching and testing approach, and the composition of our matched sets.

\subsection*{Study Population}
We use data from the NSYR, a nationally-representative survey following American youth from adolescence into early-adulthood.
The NSYR conducted four survey waves between 2002-03 (Wave 1), when subjects were between 13 and 17 years old, and 2013 (Wave 4), when subjects were 23--28 years old.
We analyze outcomes collected in Wave 4 and control for covariates measured in Wave 1.
Out of the initial 3,730 Wave 1 subjects, $2{,}144\ (57.4\%)$ were included in Wave 4.
We excluded subjects with missing primary outcomes, leaving an analytic sample of $n = 2{,}088$ subjects. 

\subsection*{Primary and secondary outcomes}

We consider two co-primary outcomes measured in Wave 4: self-reported health and the total score on the Patient Health Questionnaire \citep[PHQ-9;][]{Kroenke2001}, a standard instrument used to assess symptoms of depression.
In the Wave 4 survey, participants were asked to rate their health as either excellent, very good, good, fair, or poor.
Following precedent from earlier works \citep{IdlerBenyamini1997,Manor2000}, we dichotomized these responses by combining excellent, very good, and good into one category (coded as 0) and fair and poor responses into another (coded as 1).\footnote{This coding reverses that used in our pre-analysis protocol \cite{kokandakar2024preanalysis}. The present reverse coding simplifies the interpretation of our causal estimand as a difference in risk of self-reporting as unhealthy.}
PHQ-9 scores range from 0 to 27, with lower scores indicating better mental health.
Our secondary outcomes include body mass index; a binary indicator of being overweight; self-rated physical well-being; binary indicators for problematic and binge-drinking; and a composite life satisfaction scale score.
We defer further details about how we constructed these outcomes to Appendix~\ref{app:outcome_construction}.

\subsection*{Defining the exposure}

The NSYR did not directly record sports participation; instead, respondents were asked to list all after-school activities in which they participated (if any).
Although it is easy to categorize these responses as ``sports'' (exposed) or ``non-sports'' (control), using such broadly defined conditions may obscure important heterogeneity in the effects of participating in different kinds of sports.
At the same time, using a narrower exposure condition like ``playing football'' discards subjects who played other sports, leading to much smaller sample sizes and power.
Ultimately, the NSYR data supports multiple definitions of exposure and control conditions, and it is not \textit{a priori} obvious which to choose.

\begin{figure}
    \centering
    \begin{subfigure}[t]{0.45\textwidth}
        \centering
        \includegraphics{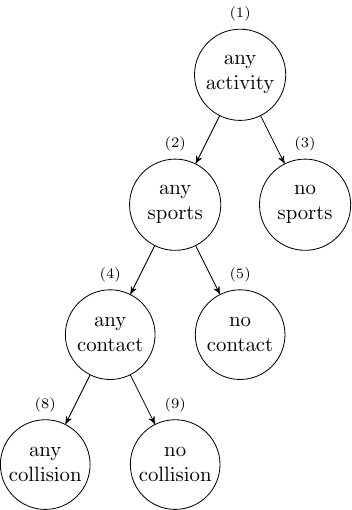}
        \caption{}
        \label{fig:trt_hierarchy}
    \end{subfigure}\hfill
    \begin{subfigure}[t]{0.55\textwidth}
        \centering
        \includegraphics{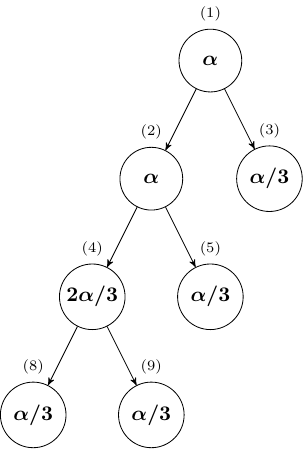}
        \caption{}
        \label{fig:trt_alphas}
    \end{subfigure}
    \caption{(a) Hierarchy of exposure definitions; (b) Type-I error rate for each hypothesis}
\end{figure}

We defined a hierarchy of exposures that retain precision in their definition while allowing sufficient sample size to power analyses that addressed our research question.
Instead of comparing ``sports'' to ``non-sports'', we first segmented our study population into increasingly finer subgroups based on their self-reported extracurricular activities.
First we divided our population into an ``any activity'' group and a ``no activity'' group.
Then, we segmented those who reported participating in any activity into a ``sports'' group and a ``non-sports'' group.
We further segmented ``sports'' into ``contact'' and ``no contact'' and finally segmented those who participated in contact sports into two more groups:  sports featuring a risk of collision and sports with no risk of collision. 
Figure~\ref{fig:trt_hierarchy} illustrates these segmentation and our definitions of contact and collision sports follow those in \cite{Meehan2016_division}.
Briefly, body-to-body and body-to-ground contact is legal and purposeful in collision sports; can occur but is not legal in non-collision contact sports; and is rare in non-contact sports; see Table~\ref{tab:sports_counts} in Appendix~\ref{app:addtl_tables}. 



While comparing the any activity group to the no activity is broadly informative about the risks and benefits of extracurricular activities, comparing the ``any collision'' group to the no activity group provides more insight into the specific benefits and risks associated with collision sports. 
However, because the any collision group is strictly smaller than the any activity group, the latter comparison is less powerful than the former. 
Rather than arbitrarily selecting one of these groups to be the exposure group, we instead conduct matched observational studies comparing each group to the ``no activity'' group, which serves as our common control group.
We conduct these matched studies and test the effect of each exposure (compared to the no activity cohort) while controlling FWER using exposure-specific Type-I error rates. 

\subsection*{Matching and randomization inference}

Letting $\tau^{(j)}$ be the \textit{constant additive exposure effect} on health of the exposure indexed by the node number $j$ in Figure~\ref{fig:trt_hierarchy}, we wish to test the null hypotheses $H_0^{(j)}: \tau^{(j)} = 0$ for each exposure group $j.$
Because our exposures are non-random, simple comparisons of the health outcomes between an exposed group and the common control are very likely to be confounded by demographic characteristics like age, gender, and race and by socioeconomic characteristics like family structure, income, and parental education. 
Table~\ref{tab:table1} summarizes the baseline covariates (all of which were measured in Wave 1) for which we control. 
To mitigate potential confounding, we would ideally first partition our sample into \emph{matched sets} containing exposed and control subjects with identical covariates and then compare the outcomes within matched sets.

Unfortunately, it is impossible to achieve this ideal in practice, and we instead create matched sets of subjects with similar baseline characteristics.
Specifically, we use optimal full matching \citep{optmatch} with a rank-based Mahalanobis distance and propensity score caliper \citep[][\S 8.3, p 171]{rosenbaum2010design}. 
The rank-based Mahalanobis distance accounts for correlation between baseline covariates, is scale-free and is insensitive to outliers. 
A propensity score caliper ensures that matched subjects have similar estimated probability of being assigned to exposure, by disallowing matching subjects whose propensity score difference is larger than a preset threshold \citep{austin2011_pscaliper}. 
We estimated a covariate-balancing propensity score \cite{ImaiRatkovic2014} using the \textbf{CBPS} package \cite{cbps_package} and built matched sets using the \textbf{optmatch} \cite{optmatch} package in \textsf{R} \cite{r_cite}.

To assess the quality of matching, we compute the standardized difference for each covariate both before and after matching.
The standardized difference for a covariate is the difference in the mean value of the covariate in the exposed and control units divided by the pooled standard error before matching.
For each covariate used in matching, we aimed for the standardized difference in means between the matched exposure and control cohorts to be in the range $(-0.2, 0.2)$ and ideally in the range $(-0.1, 0.1)$ \citep{jain_mortality_2023, rosenbaum1985constructing, cochran1973controlling}.
For each exposure, we first attempted to construct matched pairs, and then iteratively constructed matched sets with 1 exposed and $k$ control subjects or 1 control and $k$ exposed subjects, increasing $k$ until we achieved adequate balance.

\begin{table}[!ht]
    \centering
    \resizebox{\textwidth}{!}{%
    \begin{tabular}{lrrrrrrrr}
      \hline
       & & \textbf{any} & \textbf{any} & \textbf{no} & \textbf{any} & \textbf{no} & \textbf{any} & \textbf{no} \\[-2pt]
      \textbf{Covariate} & \textbf{control} & \textbf{activity} & \textbf{sports} & \textbf{sports} & \textbf{contact} & \textbf{contact} & \textbf{collision} & \textbf{collision} \\[3pt] 
      \hline \\[2pt]
    n & 573 & 1515 & 921 & 594 & 757 & 164 & 303 & 454 \\ 
      age & 15.60 & 15.48 & 15.34 & 15.69 & 15.25 & 15.78 & 15.30 & 15.22 \\ 
      gender & 51.66 & 53.47 & 45.06 & 66.50 & 37.91 & 78.05 & 9.24 & 57.05 \\ 
      \textbf{Race} & & & & & & & & \\
      \hspace{0.5em}  White (\%) & 66.67 & 75.91 & 77.20 & 73.91 & 76.22 & 81.71 & 76.57 & 75.99 \\ 
      \hspace{0.5em}  Black (\%) & 12.74 & 11.35 & 10.42 & 12.79 & 10.44 & 10.37 & 11.88 & 9.47 \\ 
      \hspace{0.5em}  Hispanic (\%) & 12.91 & 7.19 & 6.84 & 7.74 & 7.27 & 4.88 & 5.94 & 8.15 \\ 
      \hspace{0.5em}  Asian (\%) & 1.75 & 1.45 & 1.19 & 1.85 & 1.45 & 0.00 & 1.65 & 1.32 \\ 
      \hspace{0.5em}  Islander (\%) & 0.52 & 0.20 & 0.11 & 0.34 & 0.13 & 0.00 & 0.00 & 0.22 \\ 
      \hspace{0.5em}  Native American (\%) & 1.75 & 1.12 & 1.09 & 1.18 & 1.19 & 0.61 & 0.99 & 1.32 \\ 
      \hspace{0.5em}  Mixed (\%) & 2.27 & 1.32 & 1.30 & 1.35 & 1.19 & 1.83 & 0.66 & 1.54 \\ 
      \hspace{0.5em}  Other (\%) & 0.52 & 0.73 & 0.98 & 0.34 & 1.06 & 0.61 & 1.65 & 0.66 \\ 
      \hspace{0.5em}  Missing (\%) & 0.87 & 0.73 & 0.87 & 0.51 & 1.06 & 0.00 & 0.66 & 1.32 \\ 
      \textbf{Region} & & & & & & & & \\
      \hspace{0.5em}  Northeast (\%) & 13.44 & 17.89 & 19.44 & 15.49 & 19.42 & 19.51 & 17.82 & 20.48 \\ 
      \hspace{0.5em}  Midwest (\%) & 22.51 & 25.68 & 27.90 & 22.22 & 26.95 & 32.32 & 29.04 & 25.55 \\ 
      \hspace{0.5em}  South (\%) & 39.79 & 37.16 & 33.55 & 42.76 & 33.69 & 32.93 & 33.66 & 33.70 \\ 
      \hspace{0.5em}  West (\%) & 24.26 & 19.27 & 19.11 & 19.53 & 19.95 & 15.24 & 19.47 & 20.26 \\ 
      \textbf{Family Income} & & & & & & & & \\
      \hspace{0.5em} 1\textsuperscript{st} quantile (\%) & 26.00 & 13.20 & 12.49 & 14.31 & 12.55 & 12.20 & 14.52 & 11.23 \\ 
      \hspace{0.5em} 2\textsuperscript{nd} quantile (\%) & 30.02 & 22.44 & 21.17 & 24.41 & 21.53 & 19.51 & 23.43 & 20.26 \\ 
      \hspace{0.5em} 3\textsuperscript{rd} quantile (\%) & 15.01 & 17.95 & 17.05 & 19.36 & 17.44 & 15.24 & 14.85 & 19.16 \\ 
      \hspace{0.5em} 4\textsuperscript{th} quantile (\%) & 16.93 & 23.70 & 24.00 & 23.23 & 23.91 & 24.39 & 24.42 & 23.57 \\ 
      \hspace{0.5em} 5\textsuperscript{th} quantile (\%) & 12.04 & 22.71 & 25.30 & 18.69 & 24.57 & 28.66 & 22.77 & 25.77 \\ 
      \textbf{Family Structure} & & & & & & & & \\
      \hspace{0.5em}  Two parent biological (\%) & 51.31 & 60.33 & 62.54 & 56.90 & 62.62 & 62.20 & 59.74 & 64.54 \\ 
      \hspace{0.5em}  Two parent nonbiological (\%) & 13.44 & 12.54 & 12.92 & 11.95 & 12.95 & 12.80 & 12.54 & 13.22 \\ 
      \hspace{0.5em} Single parent/Other (\%) & 35.25 & 27.13 & 24.54 & 31.14 & 24.44 & 25.00 & 27.72 & 22.25 \\ 
      \textbf{Max. Parent education} & & & & & & & & \\
      \hspace{0.5em} AA/vocational degree & 20.07 & 16.37 & 15.64 & 17.51 & 15.85 & 14.63 & 15.51 & 16.08 \\ 
      \hspace{0.5em} BA/BS degree & 19.20 & 27.59 & 27.14 & 28.28 & 27.08 & 27.44 & 28.71 & 25.99 \\ 
      \hspace{0.5em}  High school degree (\%) & 42.41 & 28.84 & 27.47 & 30.98 & 28.40 & 23.17 & 30.03 & 27.31 \\ 
      \hspace{0.5em}  Higher degree (\%) & 11.87 & 24.75 & 27.25 & 20.88 & 26.16 & 32.32 & 24.09 & 27.53 \\ 
      \hspace{0.5em}  Less than high school (\%) & 6.46 & 2.05 & 2.17 & 1.85 & 2.11 & 2.44 & 1.65 & 2.42 \\ 
      \hspace{0.5em}  Missing (\%) & 0.00 & 0.40 & 0.33 & 0.51 & 0.40 & 0.00 & 0.00 & 0.66 \\ 
      \textbf{School} & & & & & & & & \\
      \hspace{0.5em}  Public school (\%) & 84.47 & 85.94 & 85.78 & 86.20 & 84.68 & 90.85 & 88.12 & 82.38 \\ 
      \hspace{0.5em}  Private school (\%) & 8.03 & 11.29 & 12.38 & 9.60 & 13.21 & 8.54 & 10.56 & 14.98 \\ 
      \hspace{0.5em}  Home schooled (\%) & 4.36 & 1.85 & 1.19 & 2.86 & 1.32 & 0.61 & 0.99 & 1.54 \\ 
      \hspace{0.5em}  Other (\%) & 3.14 & 0.73 & 0.43 & 1.18 & 0.53 & 0.00 & 0.00 & 0.88 \\ 
      \hspace{0.5em}  Missing (\%) & 0.00 & 0.20 & 0.22 & 0.17 & 0.26 & 0.00 & 0.33 & 0.22 \\[5pt] 
       \hline
    \end{tabular}}
    \caption{Summary of pre-exposure covariates before matching for each exposure}
    \label{tab:table1}
    \end{table}

\subsection*{Attrition analysis for outcomes}
We perform a separate attrition analysis for each exposure to check if the availability of both primary outcomes differs between the exposure and control cohorts.
For each exposure, we model the joint outcome availability indicator using logistic regression with the exposure indicator and pre-exposure covariates as the explanatory variables.

\subsection*{Tree-based hypothesis testing}
\label{sec:testing}
Formally, we wish to test several null hypotheses $H_{0}^{(j)}: \tau^{(j)} = 0,$ one for each of the seven exposure group (i.e., node $j$ in Figure~\ref{fig:trt_hierarchy}), which introduces a multiple comparisons problem.
To control FWER at level $\alpha = 0.05$, it is tempting to use a Bonferonni correction and test each hypothesis at $\alpha/7.$
Unfortunately, because our matched sample sizes diminish as the exposure becomes more specific, such correction can be extremely conservative.
Further, the Bonferonni correction ignores the logical relationships between hypotheses.
We instead deploy a tree-based testing-in-order strategy that leverages these relationships and permits us to test $H_{0}^{(j)}$ at level larger than $\alpha/7.$

To illustrate the basic idea, consider only the hypotheses $H_{0}^{(1)}, H_{0}^{(2)},$ and $H_{0}^{(3)},$ corresponding to comparing the no activity group to each of the any activity, any sport, and no sport groups.
If participating in any activity truly has an effect on health (i.e., $\tau^{(1)} \neq 0$), then either sports activities or non-sports activities must have some effect compared to no activity (i.e., at least one of $\tau^{(2)}$ and $\tau^{(3)}$ is non-zero).
That is, if $H_{0}^{(1)}$ is false, at most one of $H_{0}^{(2)}$ and $H_{0}^{(3)}$ can be true.
Proposition 1 of \cite{Rosenbaum_testing} shows that the sequential procedure that (i) tests $H_{0}^{(1)}$ at level $\alpha$ and then (ii) tests both $H_{0}^{(2)}$ and $H_{0}^{(3)}$ at level $\alpha$ \emph{only if $H_{0}^{(1)}$ is rejected} controls the FWER at level $\alpha.$ 


We can extend this logic to the full hierarchy of hypotheses in Figure~\ref{fig:trt_hierarchy}.
Our procedure begins by testing $H_{0}^{(1)}$ and testing each subsequent hypothesis only if its parent is tested and rejected.
If we fail to reject a null hypothesis along a branch in the tree, we do not test any of its descendents.
The procedure stops when we have either reached a non-rejection in every branch or all hypotheses have been tested.
Figure~\ref{fig:trt_alphas} shows significance levels at which we test each $H_{0}^{(j)}$ that guarantees the overall FWER is controlled at $\alpha$; see Appendix E of \cite{kokandakar2024preanalysis} for a derivation.
This tree-based testing-in-order strategy is powered to detect small effects of broad exposures and large effects of narrow exposures \cite[][\S5]{kokandakar2024preanalysis}.
Since we have two co-primary outcomes, we set $\alpha = 2.5\%$ for each of them, yielding a $5\%$ overall FWER considering both primary outcomes.
We use randomization inference for the individual tests.
For the continuous outcomes, we use the null hypothesis of no constant additive effect using the m-test implemented by the \texttt{senfm} function in the \textbf{sensitivityfull} \textsf{R} package \cite{rosenbaum_senfm}.
For binary outcomes, we test the composite null hypothesis of no difference in proportions described in \cite{fogarty_testing}.

\section{Results}
\label{sec:results}

\subsection*{Attrition Analysis}
There were $3{,}370$ respondents in Wave 1, of which only $2{,}144$ respondents were available in Wave 4.
Among them, $2{,}088$ respondents had measurements for both co-primary outcomes, i.e. there were $2{,}088$ subjects with both outcomes available (coded as 1) and $1{,}282$ subjects for whom at least one of the two primary outcomes was missing (coded as 0).
Thus, the lack of outcome availability was primarily to dropout between Wave 1 and Wave 4 of the survey.
Subjects in each exposure group were somewhat more likely to have both primary outcomes available for analysis as compared to the control group after adjusting for baseline covariates using logistic regression (See Table~\ref{tab:attr}).

\begin{table}[ht]
\centering
\caption{Odds-ratio (95\% confidence interval) and p-values for outcome availability after adjusting for potential confounders.}
\label{tab:attr}
\begin{tabular}{lccl}
  \hline
Exposure  & Odds-ratio (95\% CI) & p-value\\ 
  \hline
  any activity & 1.29 (1.10, 1.52) & 0.002  \\ 
  any sports & 1.20 (1.00, 1.44)  & 0.048 \\ 
  no sports & 1.46 (1.19, 1.80)  & 0.000 \\ 
  any contact & 1.20 (0.99, 1.45) & 0.059  \\ 
  no contact & 1.14 (0.81, 1.62)  & 0.441 \\ 
  any collision & 1.33 (1.03, 1.72)  & 0.032 \\ 
  no collision & 1.13 (0.91, 1.41) & 0.273 \\ 
   \hline
\end{tabular}
\end{table}

\subsection*{Matching results}
\begin{figure}[!h]
    \centering
    \includegraphics{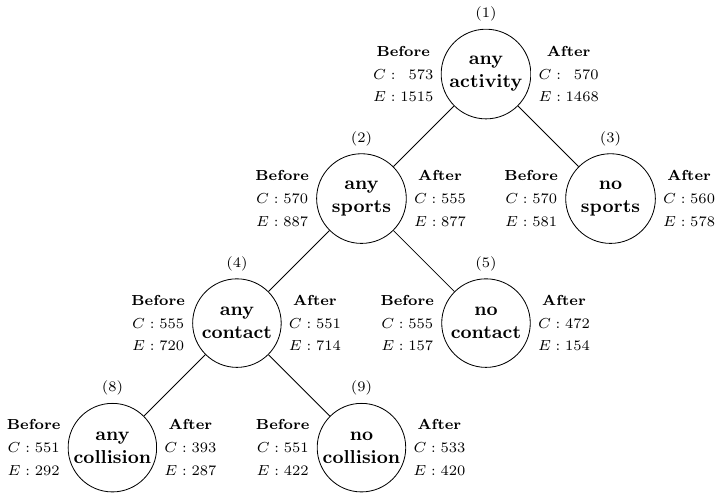}
    \caption{Number of available exposed (E) and control (C) units before and after matching for each exposure}
    \label{fig:matchtree}
    \end{figure}

Figure~\ref{fig:matchtree} shows the sizes of the exposure and control cohort both before and after matching, for each exposure.
The size of the matched exposure and control cohorts reduces as we go down the exposure hierarchy.
The matched control cohorts shrink by a larger amount than the matched exposure cohorts.
The matching procedure succeeded for each exposure i.e., all the matched covariates had a standardized mean difference of less than 0.2 units between the matched exposure and control groups for every exposure.
Moreover, there were only two instances where the standardized mean difference for a covariate exceeded 0.1 units.
In the matching process for the any activity exposure, the standardized difference for ``Northeast region'' indicator was 0.1032 which is very close to the 0.1 threshold, and therefore can be ignored.
In the second instance, the ``other race'' indicator had a standardized mean difference of 0.17 between the matched exposure and control cohorts for the no contact exposure. 
Since only four units in this category were available for matching --- a very small subset of the available units --- we are not concerned with the small bias that may be incurred due to this lack of balance.

\subsection*{Primary outcomes}
Our testing-in-order strategy reached every hypothesis for self-rated health and depression score.
Table~\ref{tab:results_comb} shows the estimates and marginal 95\% confidence intervals for the effect of each exposure on the binary self-rated health indicator. 
All the exposure cohorts indicated better self-reported health (i.e., lower risk) as compared to the control cohort consisting of those who did not participate in any organized after-school activity. 
Moreover, the any activity, any sports, and any contact sports exposures had a statistically significant and beneficial effect on self-reported health.
The risk difference between the exposure and the control groups are between 5 and 8 percentage points, with all exposure cohorts having lower risk as compared to the control group.

\begin{table}[ht]
\centering
\caption{Effect estimates (marginal 95\% confidence intervals) and p-values for the hypothesis test for each exposure. The p-values in bold denote those that are lower than the significance level in Figure~\ref{fig:trt_alphas}. For both outcomes negative differences denote better health in the exposure cohort.} 
\label{tab:results_comb}
\begin{tabular}{l@{\hspace{2em}}cl@{\hspace{2em}}cl}
  \hline
   & \multicolumn{2}{c}{Self-reported health} & \multicolumn{2}{c}{Depression Score}\\
  \cline{2-5}
  Exposure & Risk Difference (95\% CI) & p-value & Score Difference (95\% CI) & p-value \\ 
  \hline
  any activity & -0.07 (-0.13, -0.01) & \textbf{0.01}  & -0.78 (-1.37, -0.18) & \textbf{0.00}   \\ 
  any sports & -0.07 (-0.13, -0.01) & \textbf{0.01} & -1.16 (-1.79, -0.51)  & \textbf{0.00}   \\ 
  no sports & -0.05 (-0.12, 0.01) & 0.08 & -0.17 (-0.94, 0.58)  &  0.61  \\ 
  any contact & -0.08 (-0.14, -0.01) & \textbf{0.01} & -1.38  (-2.07, -0.72)  & \textbf{0.00}   \\ 
  no contact & -0.05 (-0.15, 0.07) & 0.35 & -1.06 (-2.15, 0.04)  &  0.03   \\ 
  any collision & -0.06 (-0.15, 0.04) & 0.17 & -1.12  (-2.10, -0.14)   &  0.01  \\ 
  no collision & -0.07 (-0.15, 0.00) & 0.03 & -1.25 (-2.04, -0.45)   & \textbf{0.00}  \\ 
   \hline
\end{tabular}
\end{table}

All exposure groups also displayed lower depression scores compared to the matched controls.
That said, the depression scores for the non-sports activity and the any collision cohort are statistically indistinguishable from those for the control cohort.
The any activity, any sports, any contact sports, and no collision sports cohorts all had significantly lower depression scores than the matched controls.


\subsection*{Secondary outcomes}
In our matched sample, subjects who participated in the any activity have a slightly higher BMI (effect estimate = $0.13$, CI = $(-0.48, 0.74)$), are less likely to be overweight (risk difference = $-0.02$, CI = $(-0.08, 0.04)$), are more likely to engage in problematic drinking (risk difference = $0.008$, CI = $(-0.04,0.06)$), and have a higher rate of binge-drinking (risk difference = $0.001$, CI = $(-0.044,0.045)$) as compared to the subjects in the no activity control cohort. 
However, none of these differences were statistically significant.
For each of these outcomes, our testing strategy terminated at the root node comparing any activity to no activity.

In contrast, we are able to test the null hypotheses for all the exposures for the life satisfaction score.
The life satisfaction score is higher for each exposure cohort when compared to the control cohort.
Moreover, the null hypothesis of no effect on life satisfaction score can be rejected for each exposure except for the ``no sports'' exposure.
That is, compared to the no activity cohort, the exposure cohorts other than the non-sports cohort had a significantly higher life satisfaction score.

\section{Discussion}
\label{sec:discussion}
Engaging in organized after-school activities during teenage years is associated with better overall health in the medium term (i.e., during young adulthood).
The association between better health and activity seems to be stronger for those who engaged in sports activities during their teenage years.
The finding that sports participation does have significant (albeit small) beneficial effect on mental and physical health, in line with prior findings \citep{eime_systematic_2013,bohr_et_al_2019}. 
For non-sports activities the association was not statistically significant.
One of the chief motivations of looking at different kinds of sports was concern that collision sports may potentially have substantial negative effects on health.
In fact, we find that collision sports show a small ---though not statistically significant--- improvement in health.
This finding is broadly in line with prior research that did not find large harmful effects of participating in collision sports like American football on several 
mental and physical health outcome \cite{gaulton_observational_2020,deshpande_et_al_2017,deshpande_et_al_2020,weiss_participation_2021,bohr_et_al_2019}. 

Our stepwise comparison approach suggests that, while engaging in after-school activities does improve mental health as evidenced by the decrease in depression scores, this is driven mostly by sports participation.
In fact, the control cohort and the non-sports activities cohort had very similar depression scores on average.
This seems to suggest that sports activities confer unique mental health benefits that are not associated (at least not in the same magnitude) with non-sports organized activities.

There are multiple pathways by which sports participation could manifest these benefits, including by promoting physical activity, providing venues for socialization, fostering teamwork, and encouraging positive identity development and self-esteem.
While it is certainly the case that many of the positive aspects of sports participation are also features of other after school activities, the present findings suggest that there is something about the synergy of features present in sports that confers these health benefits.

Our study has strengths and limitations.
The greatest strength is being able to determine which subgroups among those who engage in after school activities drive the statistically significant differences in health as compared to the control cohort (i.e., those who engage in no activity).
This is enabled by exploiting the logical relationships between the null hypotheses, which allows us to test each exposure effect at a higher Type I error threshold as compared to the Bonferroni correction.

One might be concerned that our conclusions may be affected by differential attrition from the NSYR.
In particular, one might worry that subjects who participated in certain sports like football may experience severe negative outcomes that lead to drop-out from the NSYR.
Under such attrition, our estimated effects would biased upwards. 
In fact, we found that subjects in each exposure group were \emph{less likely} to drop out of the NSYR between Waves 1 and 4 than the controls.
If attrition in the control group was due to experiencing extremely harmful health outcomes, then our estimated effects are biased downwards, which only amplifies the benefits of sports participation suggested by our analyses.


Throughout the study, we compare the different exposure cohorts to the common control cohort consisting of students who do not take part in after school activities.
However, it is not possible to compare two different exposure cohorts using our method, while controlling for FWER.
For instance, we cannot test if there is any effect of playing contact sports as compared to  non-contact sports.
We are able to obtain a higher Type I error threshold for each individual test as compared to a Bonferroni correction due to logical hypothesis among the hypotheses. 
However, such logical relationships may not exist for hypotheses related to comparisons between two exposure groups.
We leave the problem of comparing the health effects between two exposure cohorts while maintaining FWER for future work.

We classified sports as contact, non-contact, and collision, due to an interest in how sports-related concussion risk might moderate the health benefits of sports-participation.
However, the sports we classified in this way may differ in other respects that are relevant to physical and emotional health.
For example, individual sports such as track, cross country, swimming, and golf, predominated the non-contact sports category, whereas the majority of athletes included in the contact and collision category played team sports. Hence, we are unable in the present study to disentangle the possible benefits and risks that are due to the contact or collision nature of a sport versus those related to playing on a team. 

We tested a sharp null hypothesis of no effect for each exposure, which may be considered restrictive.
In \cite{Caughey_et_al_2023} the authors show that randomization inference can indeed be used to test for bounded null hypotheses beyond just the sharp null.
Whether their analysis extends to our tree based testing procedure remains to be seen, and we hope to explore this in future work.

Finally, because we did not have data on the incidence of sports-related concussion in this cohort, we could not evaluate whether outcomes related to sports participation are different for those who have sustained a brain injury. While our findings speak to the impact of adolescent sports participation on adult health in the aggregate, there may be subgroups of individuals for whom the risk/benefit profile differs. Understanding the contributions of concussion history and other potential risk factors is an important area for future research.  

In conclusion, we find that participating in after-school activities does have health benefits.
When compared to participating in no activity, sports participation and more specifically contact sports confer health benefits.
Therefore, our study lends credence to the idea that sports participation is a net benefit for adolescents.
In future work, we would like to extend our tree based testing procedure to enable cross-exposure group comparisons while controlling the family wise error rate.

\bibliographystyle{plainnat}
\bibliography{outcome_paper_refs,news_citations}

\clearpage
\appendix
\renewcommand{\theequation}{\thesection\arabic{equation}}  
\renewcommand{\thefigure}{\thesection\arabic{figure}}  
\renewcommand{\thetable}{\thesection\arabic{table}} 

\setcounter{equation}{0}
\setcounter{section}{0}
\setcounter{figure}{0}
\setcounter{table}{0}

\section{Outcome construction}
\label{app:outcome_construction}
\textbf{Self-rated health:}
Self-reported health is a binary indicator obtained by dichotomizing the subjects' response to the survey question ``Overall, would you say your health is: Excellent, Very good, Good, Fair, or Poor?''
Our dichotomization combines the top three responses (excellent, very good, and good) into one category (coded as 1 and interpreted as ``healthy'') and the bottom two responses (fair and poor) into another category (coded as 0 and interpreted as ``unhealthy'').
This outcome was constructed using the NSYR variable \texttt{health\_w4}.


\textbf{PHQ-9 score:}
The PHQ-9 is a standard survey instrument used to assess symptoms of depression \citep{Kroenke2001} based on answers to questions related to nine different problems like feeling tired, having little energy, or feeling down, depressed, or hopeless.
Respondents are asked how often they experience these symptoms and are given the choices ``nearly every day'', ``more than half the days'', ``several days'', or ``not at all''.
These responses are recorded in the NSYR variables \texttt{symptom1\_1\_w4}, \texttt{symptom1\_2\_w4}, \dots, \texttt{symptom1\_9\_w4}.
These responses are scored from 0 to 3 and a total score from 0 to 27 is computed for each respondent, with lower scores indicating better mental health.


\textbf{Body mass index:}
We use the body mass index (BMI) is a numeric variable recorded in the NSYR variable \texttt{bmi\_w4}, and is computed as the ratio of respondent's weight (in lbs) to the square of their height (in inches) multiplied by $703$.

\textbf{Overweight indicator:}
The NSYR variable \texttt{bmicat\_w4} records whether the respondent is underweight, normal weight, overweight, or obese based on NIH guidelines.
We convert this to a binary indicator by combining the categories overweight and obese (coded as 1), and the other two categories normal weight and underweight (coded as 0).

\textbf{Problematic Drinking:}
We use the CAGE score \citep{cage_alcohol} to construct a binary indicator of problematic drinking.
The CAGE score is constructed from four questions that were asked in the NSYR survey, ``Within the past 12 months have you ever'': (1) ``felt that you ought to cut down on your drinking?'' (\texttt{drinkcutdown\_w4}), (2) ``been annoyed with people criticizing your drinking?'' (\texttt{drinkcriticize\_w4}), (3) ``felt bad about your drinking?'' (\texttt{drinkfeltbad\_w4}), and (4) ``had a drink in the morning to steady your nerves or get rid of a hangover?'' (drinkhangover\_w4).
Each question gets a score of $0$ for a ``No'' response, and $1$ for a ``Yes'' response, adding up to a CAGE score of 0 through 4.
Note that we automatically assign a CAGE score of $0$ for subjects who indicated they never drink. 
A subject is defined as engaging in problematic drinking if the CAGE score is greater than or equal to $2$ \citep{cage_binary_threshold}.

\textbf{Binge-drinking:}
We define binge-drinking as five or more episodes drinking at least drinks in one night for females or at least 5 drinks in one night for males in a two-week period.
We constructed a binary binge-drinking indicator based on the NSYR variable \texttt{drunk\_w4}.
Additionally, subjects who indicated that they never drink (\texttt{drink\_w4}, not the same as the aforementioned binge-drinking variable) were automatically classified as not binge-drinking.

\textbf{Life Satisfaction Score:}
Similar to the scoring scheme in \cite{culver_life_sat_score} we assign a score in the range 0 (strongly disagree) to 3 (strongly agree) for each of the following survey items: (1) ``The conditions of your life are excellent'', (2) ``You are satisfied with your life'', (3) ``So far you have gotten the important things you want in life'' and (4) ``In most ways your life is close to ideal'' \citep{magolas_life_sat}.
The life satisfaction score was then constructed as the sum of the individual scores, and ranges from 0 to 12, with higher scores indicating better life satisfaction.

\section{Additional Tables}
\label{app:addtl_tables}
\begin{table}[h]
    \centering
    \small
    \caption{Sports included in our dataset, with counts in parentheses.}
    \label{tab:sports_counts}
    \begin{tabular}{lll}
        \hline
        \textbf{Non-contact} &   \textbf{Contact} & \textbf{Collision} \\
        \hline
        Track (217) &   Basketball (509)   & Football (367) \\
        Volleyball (135) &   Soccer (263)  & Wrestling (80)\\
        Cross Country (70) &  Baseball (185) & Martial Arts (37)\\
        Tennis (64) &  Softball (120)    & Lacrosse (32)\\
        Swimming (59) &   Gymnastics (22) & Hockey (30)\\
        Golf (47) &   Field Hockey (16) & Boxing (6)\\
        Racquetball (3) &  Fencing (4)   & Diving (6)\\
        Crew (9) &  Flag Football (4)   & Rugby (1)\\
        &  Water Polo (3)  &  \\
        &   Roller Hockey (2)  & \\  
        \hline  
    \end{tabular}
\end{table}

\end{document}